\documentclass[a4paper]{article}
\usepackage{subcaption}
\pdfoutput=1 
\usepackage{INTERSPEECH2020}
\usepackage{makecell}
\usepackage{floatrow}%
\usepackage{graphicx}
\usepackage[skip=3pt]{caption} 
\usepackage{tipa}
\usepackage{multirow}
\floatstyle{plaintop}
\restylefloat{table}
\title{CUCHILD: A Large-Scale Cantonese Corpus of Child Speech\\for Phonology and Articulation Assessment}
\name{Si-Ioi Ng$^{1\dagger}$, Cymie Wing-Yee Ng$^{2\dagger}$, Jiarui Wang $^1$, Tan Lee$^1$ \\ Kathy Yuet-Sheung Lee$^2$, Michael Chi-Fai Tong$^2$ \thanks{$\dagger$ Equal contribution}}
\address{
  $^1$Department of Electronic Engineering, The Chinese University of Hong Kong \\
  $^2$Department of Otorhinolaryngology, Head \&  Neck Surgery, The Chinese University of Hong Kong
  }
  
 \email{\{siioing, cymieng, jiaruiwang\}@link.cuhk.edu.hk, tanlee@ee.cuhk.edu.hk \\ \{leeys, mtong\}@ent.cuhk.edu.hk}

\begin{document}

\maketitle

\begin{abstract}
This paper describes the design and development of CUCHILD, a large-scale Cantonese corpus of child speech. The corpus contains spoken words collected from 1,986 child speakers aged from $3$ to $6$ years old. The speech materials include $130$ words of $1$ to $4$ syllables in length. The speakers cover both typically developing (TD) children and children with speech disorder. The intended use of the corpus is to support scientific and clinical research, as well as technology development related to child speech assessment. The design of the corpus, including selection of words, participants recruitment, data acquisition process, and data pre-processing are described in detail. The results of acoustical analysis are presented to illustrate the properties of child speech. Potential applications of the corpus in automatic speech recognition, phonological error detection and speaker diarization are also discussed. 
\end{abstract}

\section{Introduction}\label{intro}
Speech is one of the most common media of human communication. The natural speech sound can be captured and recorded in the form of acoustic signal for subsequent analysis. The recorded speech data are stored in a structured database that is known as a speech corpus. The speech data contain the information about the acoustic properties of speech, linguistic usage of the  language concerned, as well as the characteristics of speakers and recording conditions. With sufficient amount of speech data, statistical analysis can be performed to investigate and understand the properties of speech from different perspectives. Statistical modeling of speech data also allows the development of a wide range of speech technologies and applications. In short, speech data play an important role in multi-disciplinary research on speech communication.

The development of a specific speech technology needs to consider the target speakers and choose a suitable speech corpus according to the nature and scope of intended applications. Nowadays main-stream speech technologies are built mainly with adult speech data. They often show significantly degraded performance on child speakers, who account for a large population in the society. The performance degradation is clearly due to the differences between adult and child speech in many aspects. While abundant resources of well-annotated adult speech data are available and continually accumulated in the public domain, speech corpora of child speech are far less common. Part of this issue comes from privacy-related concerns of the parents and the difficulties in data collection due to limited attention span of child subjects.
Despite the challenges, a number of child speech corpora were developed over the years. Examples are the OGI Kids Speech corpus \cite{shobaki2000ogi}, the University of Colorado's Kids' Speech Corpus \cite{cole2006university} and the CID children's speech corpus \cite{lee1999acoustics}. These corpora target at healthy child/adolescent speakers whose ages range from $5;0$ to $17;11$. They are useful resources to support both speech technology development  \cite{yeung2018difficulties} and acoustical analysis \cite{lee1999acoustics}.  

In the population studies of speech acquisition, children of younger age tend to commit more mistakes in producing target words \cite{to2013population}. The mistakes are caused by their underdeveloped vocal tract and motor skill to produce speech sounds as well as the developing phonological abilities. 
This implies that, when a large-scale collection of child speech data is carried out, there is a high chance that speech errors would be included. By incorporating erroneous speech in the design of corpus with detailed annotation on the relevant errors, it opens a new way for error analysis and development of new systems targeting on problems of child speech acquisition. There were a few related research works on corpus development in recent years, e.g.,  \cite{kothalkar2018fusing}\cite{ramteke2019nitk}. A large part of these corpora are free from speech errors and could be used to support general speech technology development.

In this paper, we present the CUCHILD child speech corpus, which is the outcome of close collaboration between the Department of Electronic Engineering and the Department of Otorhinolaryngology, Head and Neck Surgery of the Chinese University of Hong Kong. The primary goal of this effort is to provide data resources to support acoustic analysis and identification of children with speech sound disorder, targeting the native Cantonese-speaking children of the age $3;0$ to $6;11$. The speech data would 
be useful to the research of automatic speech recognition, speaker diarization and other speech technologies.

The rest of the paper is organised as follows. Section \ref{corpus} introduces the background of Hong Kong Cantonese and describes the details of the CUCHILD corpus design. It is followed by Section \ref{analysis}, which describes the results of spectral and duration analysis of Cantonese vowels produced by children. Section \ref{applications} discusses the potential applications of the CUCHILD, followed by short conclusion in Section \ref{conclusions}.

\vspace{-3mm}
\section{Design of Corpus}\label{corpus}


\subsection{Hong Kong Cantonese}
Cantonese, which is a traditional prestige variety of the Yue Chinese Dialect group, is a major Chinese dialect widely spoken by about 68 million native speakers in Hong Kong, Macau, Guangdong and Guangxi Provinces of Mainland China, as well as overseas Chinese communities. It is a monosyllabic and tonal language. Each Chinese character is pronounced as a single syllable carrying a lexical tone. A Cantonese syllable can be divided into an onset and a rime. The onset is a consonant while the rime can contain a nucleus or a nucleus followed by a coda. The nucleus can be a vowel or a diphthong and the coda is a final consonant.  There are $19$ initial consonants, $11$ vowels, $11$ diphthongs, $6$ final consonants and $6$ distinct lexical tones (plus $3$ allotones). 
The tones are characterised by different pitches and duration patterns. 
Present-day Cantonese uses over $700$ legitimate base syllables. If tone difference is taken into account, the number of distinct syllables exceeds $1,600$ \cite{bauer2011modern}\cite{lee2002spoken}. 
\begin{table}[t!]
\centering
\caption{Number of participants in different age groups.}
\resizebox{\linewidth}{!}{%
\begin{tabular}{c|cccc}
\hline
\hline
Age (years;months)   & 3;0-3;11  & 4;0-4;11 & 5;0-5;11 & 6;0-6;11\\
\hline
Male    & $227$ & $300$ & $340$ & $113$ \\
Female    & $202$ & $368$ & $341$ & $95$ \\
\hline
\hline
\end{tabular}%
}
\label{tab:participants}
\vspace{-2mm}
\end{table}

\begin{table}[t!]
\centering
\caption{Distribution of participants/ kindergartens in different districts.}
\resizebox{\linewidth}{!}{%
\begin{tabular}{c|cccc}
\hline
\hline
Districts  & Hong Kong Island  & New Territories & Kowloon\\
\hline
Kindergartens    & $7$ & $7$ & $3$ & \\
Participants    & $740$ & $761$ & $485$ & \\
\hline
\hline
\end{tabular}%
}
\label{tab:speaker_distribution_districts}
\vspace{-2mm}
\end{table}

\subsection{Participants}
The speech samples in the CUCHILD corpus were collected from 1,986 Hong Kong pre-school children (1,006 female, 980 male, age 3;0 to 6;11)  during the period from February 2017 to January 2018. All speakers use Hong Kong Cantonese as their first language (L1). The children were grades K1 to K3 students, recruited via normal local kindergartens which use Cantonese as their medium of teaching. Children from the special child care centres were not included. Parental consents were obtained for each participating child. Information on age and gender were collected and they are summarized in Table \ref{tab:participants}. 17 kindergartens from different districts of Hong Kong participated in the study and the information of their distribution is presented in Table \ref{tab:speaker_distribution_districts}.

\subsection{Recording session setup}
Each participant was seen individually in a separated area inside the kindergarten. He/she was arranged to sit face-to-face in front of a research assistant with a mini-game setting to engage his/her attention. 
A digital recorder (TASCOM DR-44WL) was located at $20$-$50$ centimeters in front of the children’s mouth. As the environmental noise such as reverberation, school bells, people walking around, etc. was unavoidable, the gain of the recorder was adjusted to maintain the background noise level below -$30$ dB (relative to the maximum input level) with the best effort.
The sampling rate was set to be $44.1$ kHz with two-channel stereo recording. 

For each child subject, the recording contains an interactive conversation between the child and the research assistant. The research assistants were student clinicians from speech therapy programmes in local universities. A technician, who was a student with engineering background, was responsible to monitor the operation of recording devices.
As children would lose concentration easily, sufficient break time was allowed during the session. With previous experience of working with children, the research assistants were able to engage the participants with the mini-game and elicit targeted verbal outputs during the sessions. The majority of the participants were co-operative in the recording process.

Each recording session consisted of two major parts involving three stimuli booklets. In the first part, a single word articulation test, namely Hong Kong Cantonese Articulation Test (HKCAT)\cite{cheung2006hong}, was used to obtain the information about the child's speech sound ability at single word level. In the second part, the subject were asked to read aloud two stimuli booklets with pictures that illustrate 130 Cantonese words (223 syllables) with 1-4 syllables.

\subsection{Composition of stimuli}
Hong Kong Cantonese Articulation Test (HKCAT) is a standardized single word articulation test commonly used by qualified speech therapists in Hong Kong. It provides information about the speech sound inventory, speech sound errors and patterns of the participant. All research assistants had received proper training on the use of HKCAT and transcription of Cantonese speech sounds. The procedure during data collection was monitored by the supervisor, who is a qualified speech therapist with more than 10 years of clinical experience in working with children with speech sound disorders. The results of the HKCAT were instantly transcribed on recording forms by the research assistants.

After HKCAT, the child subject was asked to name the 130 Cantonese words one by one. When a subject failed to name a picture, the research assistant would provide a direct model for the child to repeat and imitate. The target words were selected with the consideration of their age-adequacy and are illustrated with children-friendly colorful drawings. Samples of the stimuli are illustrated in Figure 1(a)-(d). These words were selected with an aim to elicit and collect speech samples covering all Cantonese phonemes in words of different lengths, with different syllable structures (CV, CVV, CVC) and at different syllable positions. The initial consonants include plosives, affricates, nasals, fricatives, approximants and lateral approximants. The initial consonant [n] was not included as it is commonly regarded as an allophone of [l] in Hong Kong Cantonese. Seven long vowels, four short vowels, eleven diphthongs, six final consonants and six lexical tones were all covered in the 223 syllables. The list of phonemes are summarized as in Table \ref{tab:phonemes_CUCHILD}.

\begin{table}[h!]
\centering
\begin{tabular}{c|c}
\hline
\hline
  & Phonemes \\
\hline
Initial consonants    &  \makecell{p p\textsuperscript{h} t t\textsuperscript{h}  k k\textsuperscript{h}  k\textsuperscript{w} k\textsuperscript{wh} \\ts ts\textsuperscript{h} m \textipa{N f s h w j l}}
\\
\hline
Long vowels    & \textipa{a: i: E: œ: O: u: y:} \\
\hline
Short vowels  & \textipa{5 I 8 U}\\
\hline
Diphthongs & \makecell{ \textipa{ai ei 5i ui Oi} \\ \textipa{au 5u iu ou 8y Eu} }
\\
\hline
Final consonants & -p -t -k -m -n -\textipa{N}\\
\hline
Tones & \makecell{High-level Mid-rising\\ Mid-level Mid-falling \\Low-rising Low-level}
\\
\hline
\hline
\end{tabular}%
\caption{Cantonese phonemes included in CUCHILD}
\label{tab:phonemes_CUCHILD}
\vspace{-2mm}
\end{table}


\begin{figure}[th!]
    \setlength\belowcaptionskip{-0.4\baselineskip}
      \centering
\begin{subfigure}[b]{0.24\textwidth}
\includegraphics[width=\textwidth]{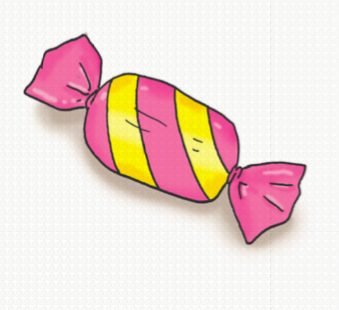}
\caption{}
\end{subfigure}
\begin{subfigure}[b]{0.24\textwidth}
\includegraphics[width=\textwidth]{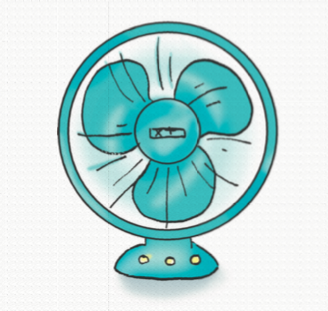}
\caption{}
\end{subfigure}
\begin{subfigure}[b]{0.24\textwidth}
\includegraphics[width=\textwidth]{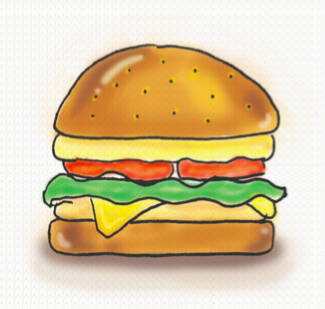}
\caption{}
\end{subfigure}
\begin{subfigure}[b]{0.24\textwidth}
\includegraphics[width=\textwidth]{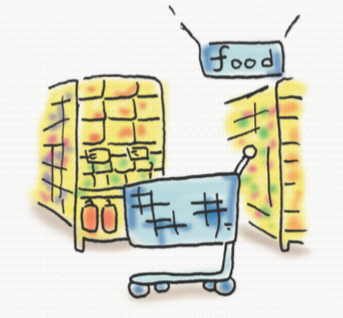}
\caption{}
\end{subfigure}
        \caption{Samples of stimuli: Cantonese words with 1-4 syllables. (a) "{t\textsuperscript{h}\textipa{O:N}25}" (Candy) (b) "{f\textipa{UN}55 s\textipa{i:}n33}"(Fan) (c) "{h\textipa{O:}n33 pou25 p\textipa{a:}u55"(Hamburger) (d) "{ts\textsuperscript{h}\textipa{i:}u55 k\textsuperscript{h}\textipa{5}p55 s\textipa{i:}33 ts\textsuperscript{h}\textipa{œ:N}21" (Supermarket)}}                 
        } 
        \label{Word_samples}
\end{figure}

\vspace{-3mm}
\subsection{Pre-processing of collected speech data}
Upon the collection of the speech samples from the children, the HKCAT results charted by the research assistants were validated by the supervisor, partially onsite and entirely at the laboratory with reference to the audio and audio-visual recordings. An analysis of the screening results from face-to-face analysis, audio recordings and audio-visual recordings suggested that no significant difference was found with the HKCAT scores among different modes of judgement \cite{ng2018comparison}. The HKCAT scores provide important information about the children's speech inventory, speech sound errors and patterns, and serve as a reference transcription of the collected speech data with the 223 syllables in CUCHILD. Age appropriate errors made by typical developing children, age inappropriate phonological processes produced by children with suspected speech sound disorders and articulation errors are included. The referenced transcription is used to categorise the collected speech data into accurate pronunciation and expected erroneous speech collected from typically developing (TD) children and unexpected erroneous speech collected from children with disordered speech (DS). Thus, in addition to the full coverage of all Cantonese phonemes, the speech data in CUCHILD give a spectrum of phonological processes and articulation errors which are typically/ atypically found in Cantonese-speaking children at age 3;0 to 6;11. 
The pre-processed information and manual annotation of speech data can be used as training data for speech recognition system 
as well as the other proposed functions and applications.

\section{Acoustical Analysis}\label{analysis}

\begin{figure}[t!]
        \setlength\belowcaptionskip{-0.25\baselineskip}
       \centering
\begin{subfigure}[b]{0.49\textwidth}
\includegraphics[width=\textwidth]{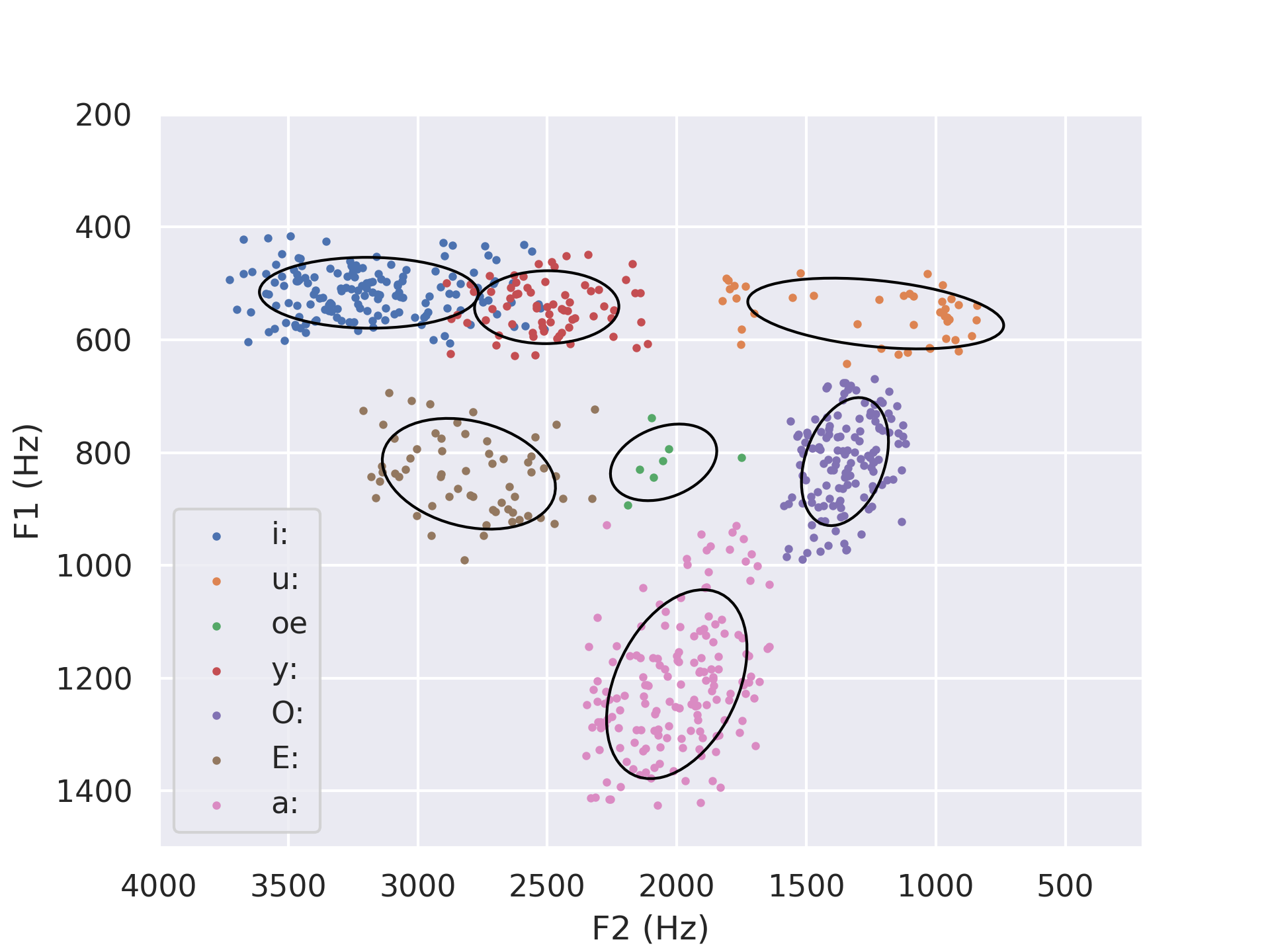}
\caption{}
\end{subfigure}
\begin{subfigure}[b]{0.49\textwidth}
\includegraphics[width=\textwidth]{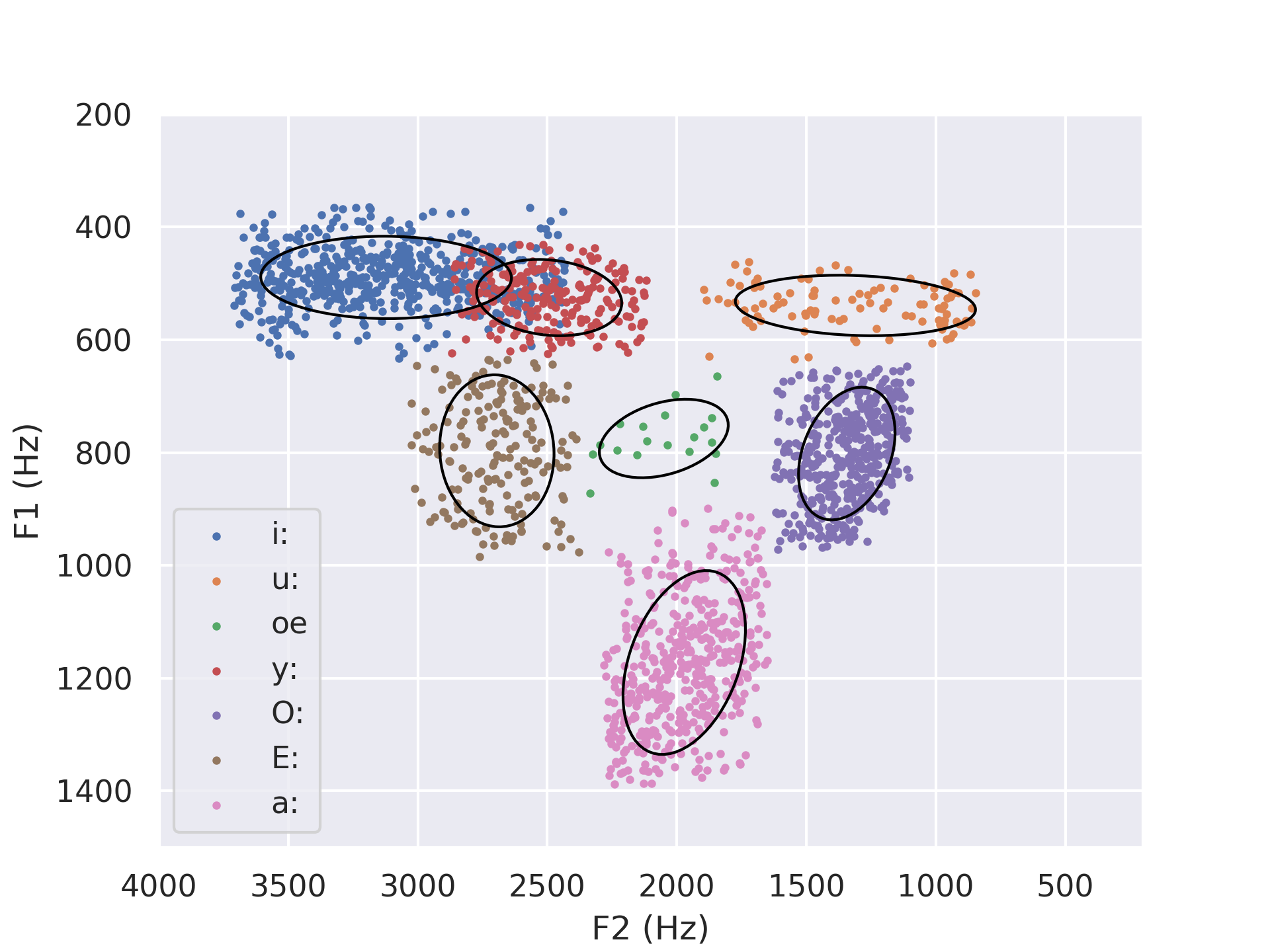}
\caption{}
\end{subfigure}
\begin{subfigure}[b]{0.49\textwidth}
\includegraphics[width=\textwidth]{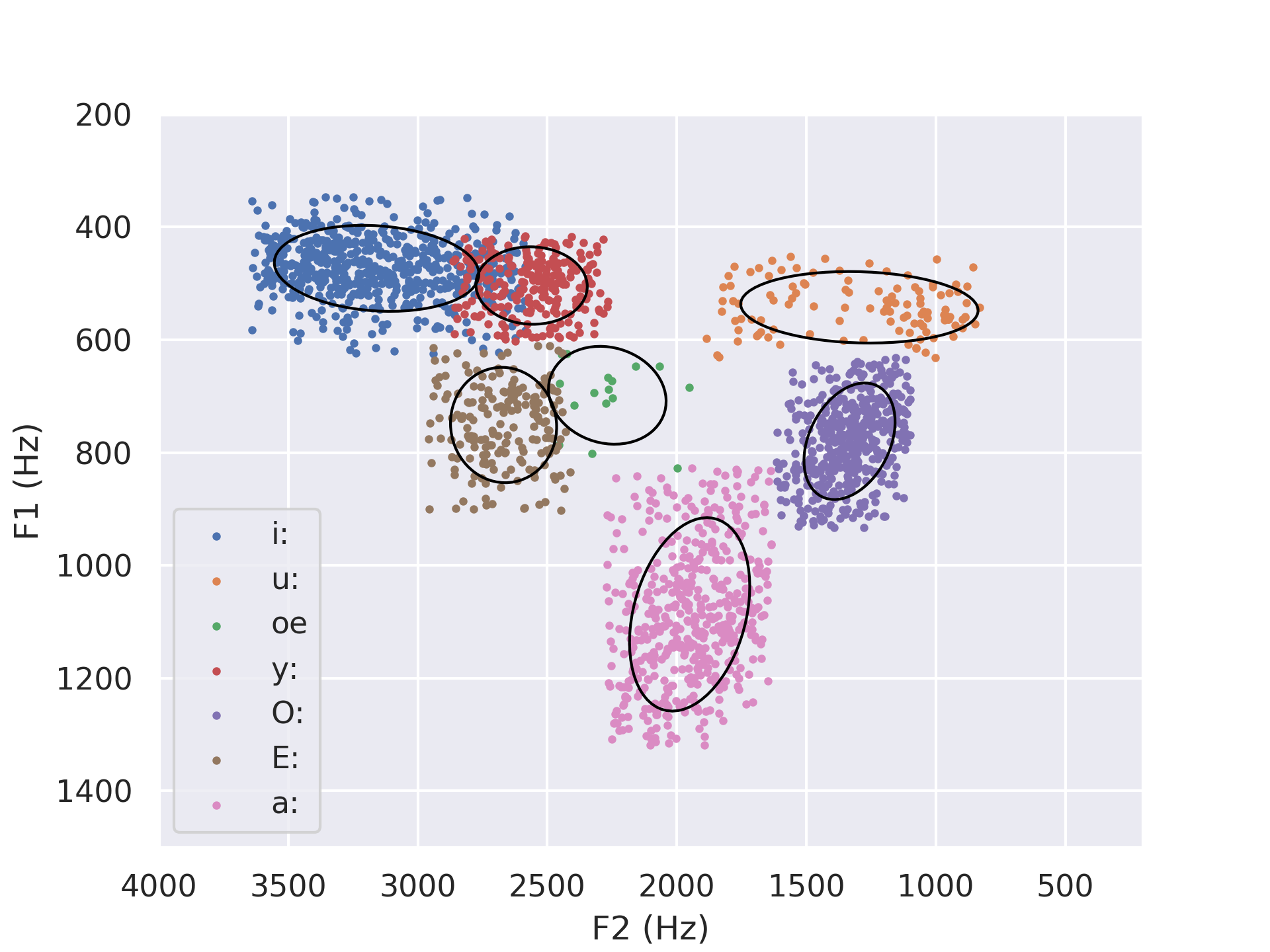}
\caption{}
\end{subfigure}
\begin{subfigure}[b]{0.49\textwidth}
\includegraphics[width=\textwidth]{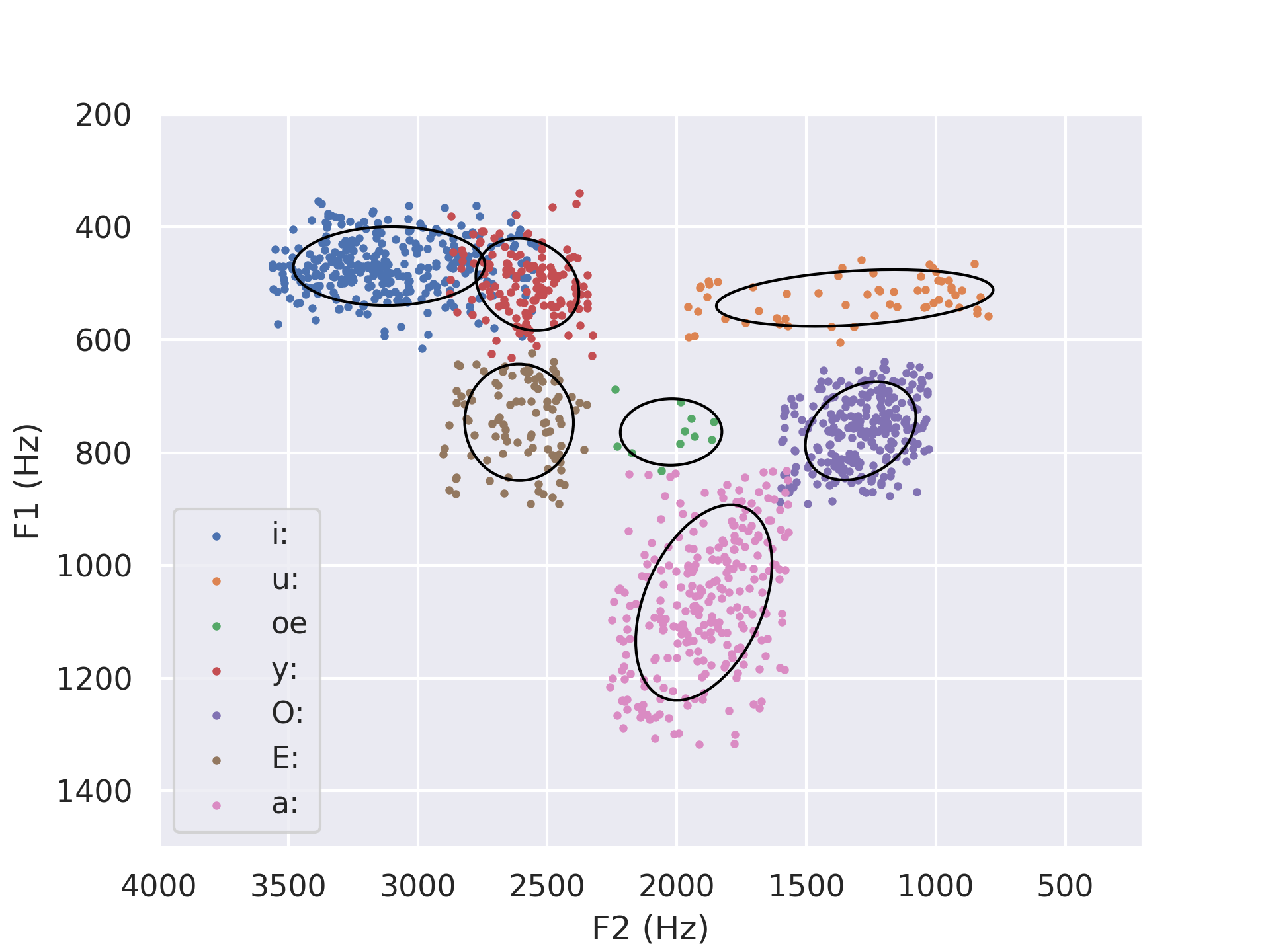}
\caption{}
\end{subfigure}
     
        \caption{Formant analysis of the $7$ Cantonese vowels over different age group, illustrated by F1-F2 scatter plots: (a) Age between $3;0$ - $3;11$; (b) Age between $4;0$ - $4;11$; (c) Age between $5;0$ - $5;11$; (d) Age between $6;0$ - $6;11$.
        The legend (from top to bottom) represents the vowels [\textipa{i: u: \oe: y: O: E: }].} 
        \label{formant_plot}
\end{figure}

Acoustical analysis of child speech aims to provide better understanding about developmental changes of acoustic patterns. Clinically the findings can provide a reference of each speaker group for assessment.
In this section, 
we measure the fundamental frequency (F0) and 
the first two formants (F1, F2)
of the $7$ Cantonese long vowels [\textipa{i: y: E: \oe: a: O: u:}] using automatic F0 and formant tracking algorithms. 
The long vowels involve $34$ monosyllable words with the syllable structure of (C)V:. The effect of lexical tone is not considered in this study.

\begin{table}[t]
\centering
\resizebox{\linewidth}{!}{%
\begin{tabular}{c|cccc}
\hline
\hline
Age (years;months)   & 3;0-3;11  & 4;0-4;11 & 5;0-5;11 & 6;0-6;11\\
\hline
Male    & $6$ & $25$ & $29$ & $12$ \\
Female    & $13$ & $32$ & $34$ & $20$ \\
\hline
\hline
\end{tabular}%
}
\caption{Number of speakers used in acoustic analysis.}
\label{tab:participants_acoustic}
\vspace{-2mm}
\end{table}

A subset of speech data is selected from $171$ TD speakers, as summarized in Table \ref{tab:participants_acoustic}. The audio signals are down-sampled from $44.1$ kHz to $16$ kHz and converted to single-channel signals. Each target word in the recording is manually segmented and transcribed by trained research assistants using the software Wavesurfer \cite{sjolander2000wavesurfer}. To locate the vowel segments for subsequent analysis, forced alignment is applied to the speech data with a GMM-HMM triphone acoustic model. The acoustic model is trained with $13$-dimensional Mel-frequency cepstral coefficients (MFCC) and their first- and second-order derivatives, which are extracted every $10$ ms with a $25$ ms Hamming window. Linear discriminant analysis (LDA), semi-tied covaraicne (STC) transform and feature space Maximum Likelihood Linear Regression (fMLLR) are also applied in the tri-phone model training \cite{duda2012pattern}\cite{gales1999semi}\cite{gales1998maximum}.
The acoustic modeling and forced alignment are implemented using the Kaldi speech recognition toolkit. \cite{povey2011kaldi}. 
Vowel segments shorter than $100$ ms are not included in the analysis. F0 and formant frequencies are estimated by Praat using the auto-correlation method and linear predictive analysis with Burg's algorithm respectively \cite{boersma1993accurate} \cite{andersen1974calculation}\cite{boersma2018praat}\cite{jadoul2018introducing}.

Child speech is known to have higher F0 and formant frequencies than adult speech. The wide spacing of harmonic peaks makes the analysis more difficult \cite{kent2018static}. To avoid erroneous estimation of formant frequencies, the ceiling values of formant frequencies for front vowels [\textipa{i: y: E: \oe:}], central vowel [\textipa{a:}] and back vowels [\textipa{O: u:}] are empirically set to be $8,000$ Hz, $7,000$ Hz and $6,000$ Hz respectively. We allow a maximum of $5$ formants (F1 - F5) to be estimated in each analysis frame. For F0 estimation, the pitch floor is set to be $120$ Hz. 

\begin{figure}[t]
    \centering
    \includegraphics[width=0.75\linewidth]{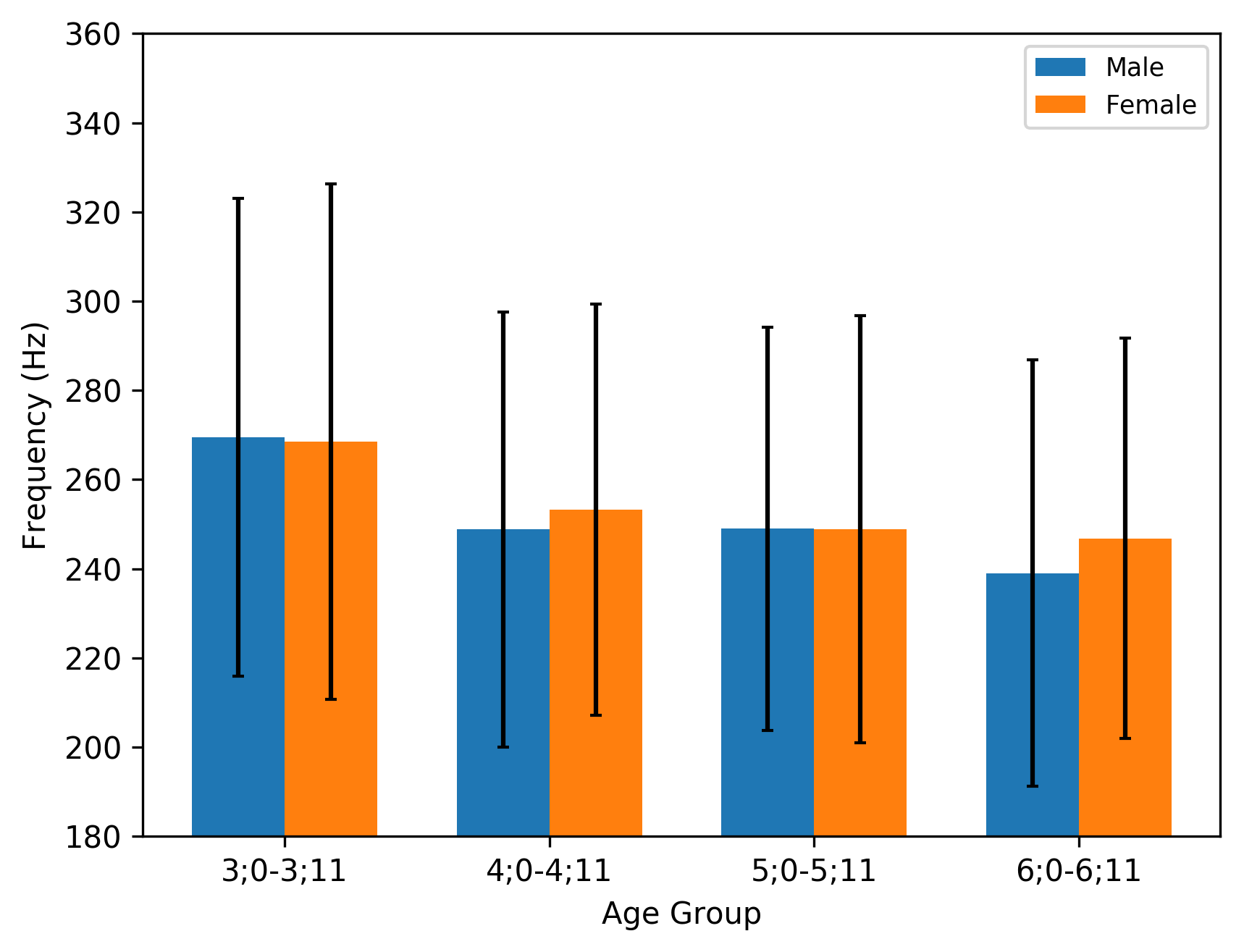}
    \caption{Results of fundamental frequency analysis of different age groups and genders} 
    \label{fig:f0_plot}
    \vspace{-3mm}
\end{figure}

Each vowel segment consists of a number of analysis frames, from each of which F0 and formant frequencies can be extracted. The median values over all frames are used to represent the whole segment. The mean F0 values of male and female speakers are listed as in Figure \ref{fig:f0_plot}. As the age increases, the child speakers of both genders show a declining trend in F0. Boys generally have lower F0 than girls, but the difference is very small. At age of $4$, boys have a mean F0 of $249$ Hz whereas the mean F0 of girls is $253$ Hz. At age of $6$, the mean F0 values of boys and girls are $239$ Hz and $247$ Hz respectively.


Estimation of formant frequencies exhibits frequent occurrences of errors, especially that closely located formants may not be identified. A procedure of data cleansing is applied to make the statistical analysis more meaningful. Estimated raw values for each formant (F1 - F3) are grouped according to vowel identity, age and gender. For each group, the mean and standard deviation are computed. Any measured value deviating by $1.440$ standard deviation from the mean is removed.
The F1-F2 plots for different age ranges are illustrated as in Figure \ref{formant_plot}. Different vowels are marked in different colors. The vowel ellipses are drawn to represent the $85$\% confidence interval. It is known that the F1 value is related closely with the height of tongue, whereas F2 is determined mainly by the frontness and backness of the tongue body. 
The mean values of F1 and F2, as well as the mean duration of $5$ vowels [\textipa{i: E: a: O: u:}] for the age groups of $3$ and $6$ are shown as in Table \ref{tab:formant_table}. 
Comparing the two age groups, there is a trend of decrease in F1 values of all $5$ vowels when the age increases. Similar observation applies to F2 except for \textipa{[u:]}. The vowel duration by children of age $3$ is slightly longer than those of age $6$.




\begin{table}[h!]
    \vspace{-3mm}
    \caption{Formant values and duration of 5 long vowels [\textipa{i: E: a: O: u:}] which are commonly used to represent the vowel loop.} 
    
    \centering
    \resizebox{\linewidth}{!}{%
     \begin{tabular}{|c|c|c|c|}
    \hline
    \multirow{2}{*}{Vowel} & \multicolumn{3}{c|}{Age $3$ / Age $6$} \\
    \cline{2-4}
      & F1 (Hz) & F2 (Hz) & Duration (s)  \\
    \hline
    [\textipa{i:}] & $511$ / $485$ & $3129$ / $3070$  & $0.26$ / $0.26$  \\
    \hline
    [\textipa{E:}] & $850$ / $755$ & $2777$ / $2630$  & $0.28$ / $0.282$  \\
    \hline
    [\textipa{a:}] & $1181$ / $1068$  & $1998$ / $1913$  & $0.31$ / $0.29$ \\
    \hline
    [\textipa{O:}] & $827$ / $768$  & $1353$ / $1314$  & $0.25$ / $0.24$ \\
    \hline
    [\textipa{u:}] & $563$ / $534$ & $1306$ / $1378$  & $0.25$ / $0.21$ \\
    \hline
    \end{tabular}
    }  
    \label{tab:formant_table}
\end{table}

\vspace{-6mm}
\section{Applications of CUCHILD}\label{applications}

\subsection{Speech recognition and speaker diarization}
In automatic speech recognition (ASR), the high diversity of acoustic properties and limited language proficiency in child speech explain that statistical models trained from adult speech are not applicable to child speech. The presence of child speech data is necessary in the development of ASR systems for child users. The CUCHILD corpus is expected to address the issue by providing a large amount of child speech data. 
Speaker diarization (SD), aiming to solve the "who speaks when" problem, is another important research topic with practical significance. Currently SD systems are commonly trained on adult speech. The spontaneity and phonetic variation in child speech make the extraction of speaker information difficult \cite{xie2019multi}. A high-performance SD system for child speech is expected to bring the benefit in different aspects. For instance, a SD system can be used to analyze adult-child interaction and extract target speech from child in a conversation \cite{kothalkar2019tagging}. The extracted child speech data can be used to provide training data for ASR system development \cite{wang2018study} or support the development of clinical assessment tools \cite{shahin2019automatic}. In addition, the analysis of adult-child speech interaction would be helpful to understanding children's typical or atypical social behaviours \cite{hansen2019speech}. 


\subsection{Detection of speech sound errors}
Speech Sound Disorder (SSD) is diagnosed when a child shows difficulties in acquisition, production and perception of speech, and makes errors in pronunciations that do not match the normal variation expectation for his/her age\cite{WinNT}. Poor speech sound production skills are 
found to have significant impacts on social, emotional and academic developments\cite{hitchcock2015social}, and associated with lower literacy outcomes \cite{overby2012preliteracy}, \cite{lewis2011literacy} and a greater likelihood of suffering reading disorders\cite{peterson2009influences}. 
With large amount of child speech data, automatic detection of phonological and articulation errors is feasible using the machine learning approach. 
Automatic detection tools are expected to accelerate the screening of children who are at-risk for SSD, thus bringing early identification and intervention. In the long-run, early intervention can bring positive impacts to the children development, and thus reduce the service load of the current healthcare system on children with special education needs. 
The CUCHILD includes recordings of accurate production and expected erroneous speech produced by TD children, as well as the unexpected erroneous speech produced by disordered children. 
It is designed to support the development and evaluation of the detection systems. Relevant works can be found in \cite{ng2018automated}\cite{wang2019child}. 


\subsection{Developmental studies on children}
Children's acquisition of speech sounds can be investigated by large scale population studies, as in \cite{to2013population}\cite{so1995acquisition}.  Using the articulation test, the subject-level statistics of the results describe the overall picture of phonological acquisition and indicate the developmental error patterns. These studies often involve huge demand in manpower and professional costs, and take long period of time in data-collection, validation and drawing result conclusion. 
Alternatively, child speech can be collected and analysed based on acoustic signal. The signal captures rich linguistic and speaker information. The findings from the studies of acoustic features can bring new insight to the developmental changes of child speech, as well as inspire new approaches to differentiate atypical from healthy speech with the automated system. The CUCHILD satisfies the above-mentioned motivations and supports the studies of acoustic properties of pre-school child speech.


\section{Conclusion}\label{conclusions}
In this paper, a large-scale child speech corpus CUCHILD with Cantonese speech sounds collected from 1,986 children of age from 3;0 to 6;11 is presented. The corpus includes the recordings of the speech sounds collected from both typically developing children and children with disordered speech when reading 130 Cantonese words with 1 to 4 syllables. All initial consonants, vowels, diphthongs, final consonants and lexical tones of Cantonese were covered in the corpus. Acoustical analysis with a subset of speech sample including the measurement of fundamental frequency and the first three formants was illustrated in this paper. Future work with the corpus includes child speech recognition, speaker diarization, detection of speech sound errors and further spectral analysis are suggested and to be investigated.   

\section{Acknowledgements}\label{ack}
This research was partially supported by a direct grant and a Research Sustainability Fund from the Research Committee of the Chinese University of Hong Kong, as well as the financial support by the Hear Talk Foundation under the project titled "Speech Analysis for Cantonese Speaking Children".  


\bibliographystyle{IEEEtran}

\bibliography{me}

\end{document}